\documentstyle[12lomcon,cite,graphicx]{article}

\begin{document}

\title{ISR PHYSICS AT BABAR}

\author{ V.Druzhinin \footnote{e-mail: druzhinin@inp.nsk.su}\\
Representing the BaBar Collaboration}

\address{Budker Institute of Nuclear Physics, 630090 Novosibirsk, Russia}

\maketitle\abstracts{ 
We present a review of BaBar results on e+e- annihilations into
exclusive hadronic final states using the initial state radiation
technique.  Cross sections over the $\sqrt{s}$ range from threshold to
4.5 GeV, with very small point-to-point systematic errors, are presented
for the $3\pi$, $2(\pi^+\pi^-)$, $3(\pi^+\pi^-)$, $2(\pi^+\pi^-)2\pi^0$,
$K^+K^-\pi^+\pi^-$, $2(K^+K^-)$ and $p\bar{p}$ final states.  The proton
form factor and the ratio of its electric and magnetic components are
also presented.}
\section{Introduction}
The BABAR detector~\cite{babar} collect data at the PEP-II asymmetric-energy 
$e^+e^-$ collider where 9-GeV electrons collide with 3.1-GeV positrons.  
The detector covers about 85\% of the solid angle in the $e^+e^-$ 
center-of-mass (c.m.) frame and provides excellent charged particles tracking 
and identification, and photon detection.

These detector features and high luminosity collected at the PEP-II allow
to study the processes of $e^+e^-$ annihilation into different hadronic states
using initial state radiation (ISR) technique. In the ISR processes, 
$e^+e^- \to \gamma X$, a photon of energy $E_{\gamma}$ is emitted by 
the initial electron or positron. As result the produced hadronic system 
has invariant mass $m=\sqrt{s(1-x)}$, where $\sqrt{s}$ is $e^+e^-$
c.m. energy and $x=2E_{\gamma}/\sqrt{s}$.  
The mass spectrum of the hadronic system $X$ is related to the 
$e^+e^- \to X$ cross section as
\begin{equation}
\frac{{d}^2\sigma(e^+e^-\to \gamma X)}{{d}x {d}\cos{\theta_\gamma}}=
W(s,x,\theta_\gamma)\sigma(m).
\end{equation}
Here $W$ is well-known photon radiator function describing ISR photon
energy and angular distribution. The ISR photon is emitted predominantly
along electron or positron direction. In our approach with detection of
the ISR photon at large angle we use only about 10\% of ISR events.
\section{Selection of ISR events and background subtraction}
For analysis we select events with all final particles (including hard
ISR photon) detected. The charged particle identification is used to
recognize the specific hadronic state and suppress background.
For selected events we perform the kinematic fit with the requirement 
of energy and momentum balance. The two-photon mass for $\pi^0$ candidates 
is constrained to $\pi^0$ nominal mass. The cut on $\chi^2$ of the 
kinematic fit provides additional background suppression.

\begin{figure}
\includegraphics[width=.35\linewidth]{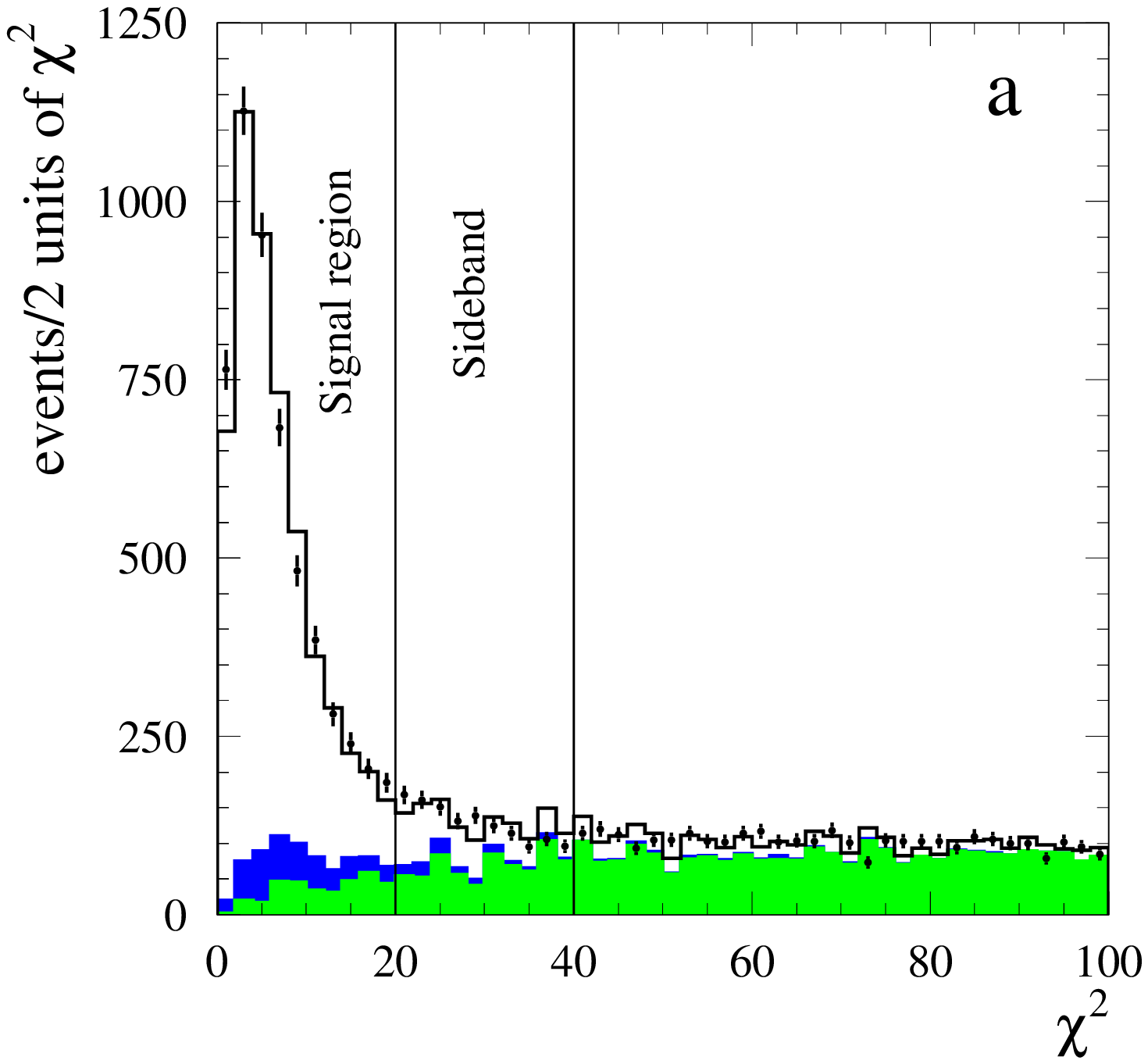}
\hfill
\includegraphics[width=.64\linewidth]{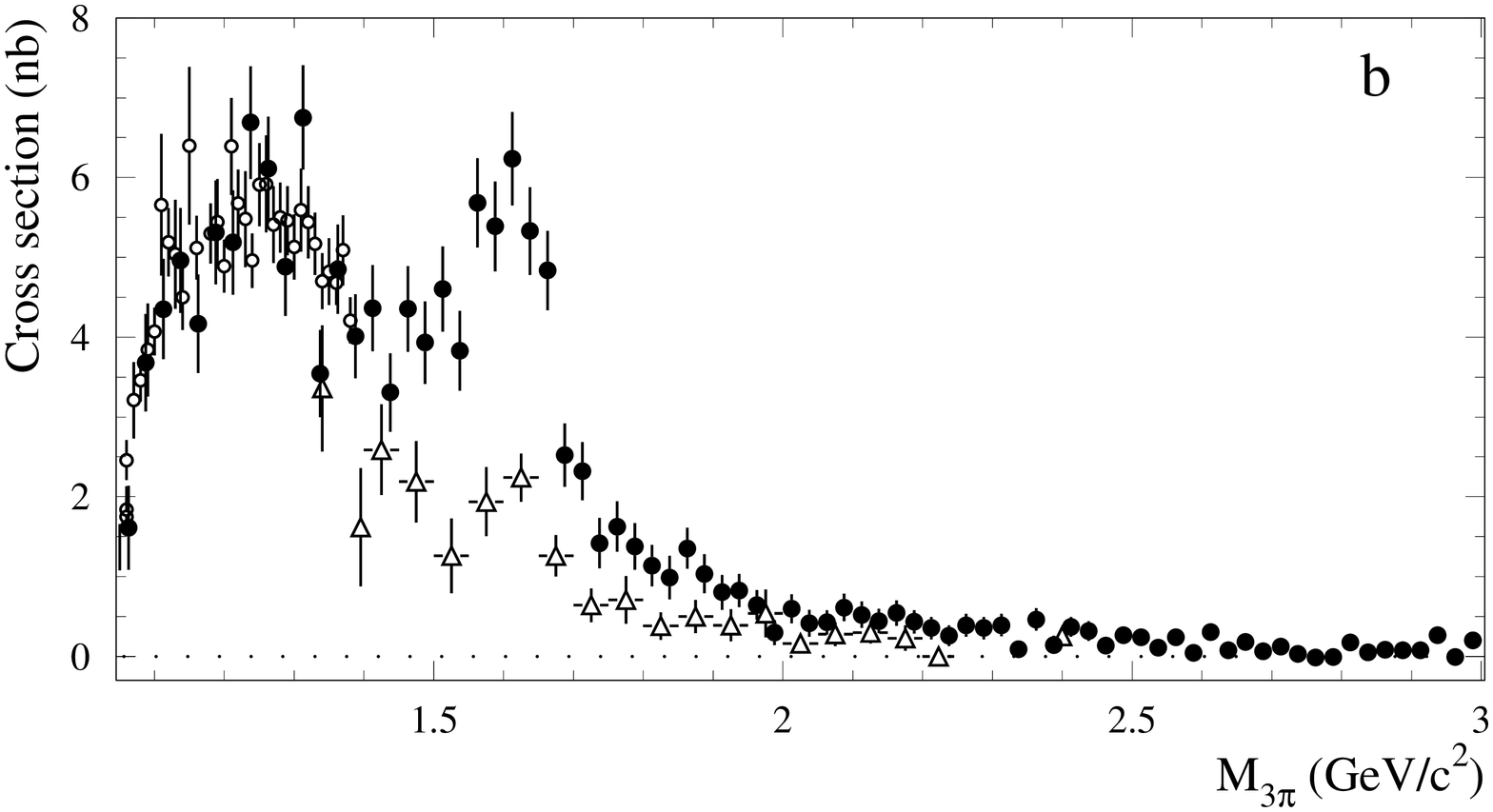}
\caption{a). The $\chi^2_p$ distribution for selected $3\pi\gamma$ candidates
for data (points with error bars) and simulation (histogram).
The dark and lightly shaded histograms shows the distributions for
$e^+e^-\to \pi^+\pi^-\pi^0\pi^0$ and other background processes, 
respectively. b). The $e^+e^-\to \pi^+\pi^-\pi^0$ cross section measured
by BABAR (filled circles), by SND (open circles), and DM2 (open triangles).
\label{fig1}}
\end{figure}

The main sources of remaining background are the ISR events with
misidentified particles ($K^+K^-\pi^0\gamma$ for $3\pi\gamma$ process,
$K^+K^-\gamma$ and $\pi^+\pi^-\gamma$ for $p\bar{p}\gamma$),
events with high-energy $\pi^0$ instead of ISR photon ($p\bar{p}\pi^0$ for
$p\bar{p}\gamma$). The third class of background processes is all other
ISR processes and processes of continuum annihilation into quark-antiquark.
For example, the $\chi^2$ distribution for $e^+e^- \to  \gamma \pi^+\pi^-\pi^0$
candidates is shown in  Fig.~\ref{fig1}a and seen to be well described by our
simulation.
Backgrounds from ISR events with misidentified particles
($K^+K^-\pi^0\gamma$) and events with a high-energy $\pi^0$ mimicing an ISR
photon ($\pi^+\pi^-\pi^0\pi^0$) also peak at low $chi^2$ values (dark shaded
histogram in Fig.~\ref{fig1}a), but are a small fraction of the signal, 
and are measured from the data as a function of $3\pi$ mass and subtracted.
Other sources of background involve missing or spurious particles and
give a broad $\chi^2$ distribution (light shaded histogram); they are
subtracted using the $\chi^2$ sideband region indicated.
\section{Results}
\underline{\bf\boldmath The $\pi^+\pi^-\pi^0$ final state}~\cite{3pi}.
The $e^+e^-\to 3\pi$ is a key process for study of
the excited $\omega$ states. The measured $e^+e^-\to 3\pi$ cross 
section for the mass above 1.05 GeV is shown in Fig.~\ref{fig1}b. 
Our results are consistent with data obtained by SND collaboration
for energies below 1.4 GeV, but significantly exceed
previous DM2 results for the mass region of $\omega^{\prime\prime}$ resonance.
From the fit of the cross section by a sum of the 
contributions of $\omega$, $\phi$, $\omega^{\prime}$, and 
$\omega^{\prime\prime}$ mesons
we obtained the parameters of excited $\omega$ states: 
$M_{\omega^{\prime}}=1350\pm30$ GeV/$c^2$,
$\Gamma_{\omega^{\prime}}=450\pm100$ GeV/$c^2$,
$M_{\omega^{\prime\prime}}=1660\pm10$ GeV/$c^2$,
$\Gamma_{\omega^{\prime\prime}}=230\pm40$ GeV/$c^2$.

\begin{figure}
\includegraphics[width=.32\linewidth]{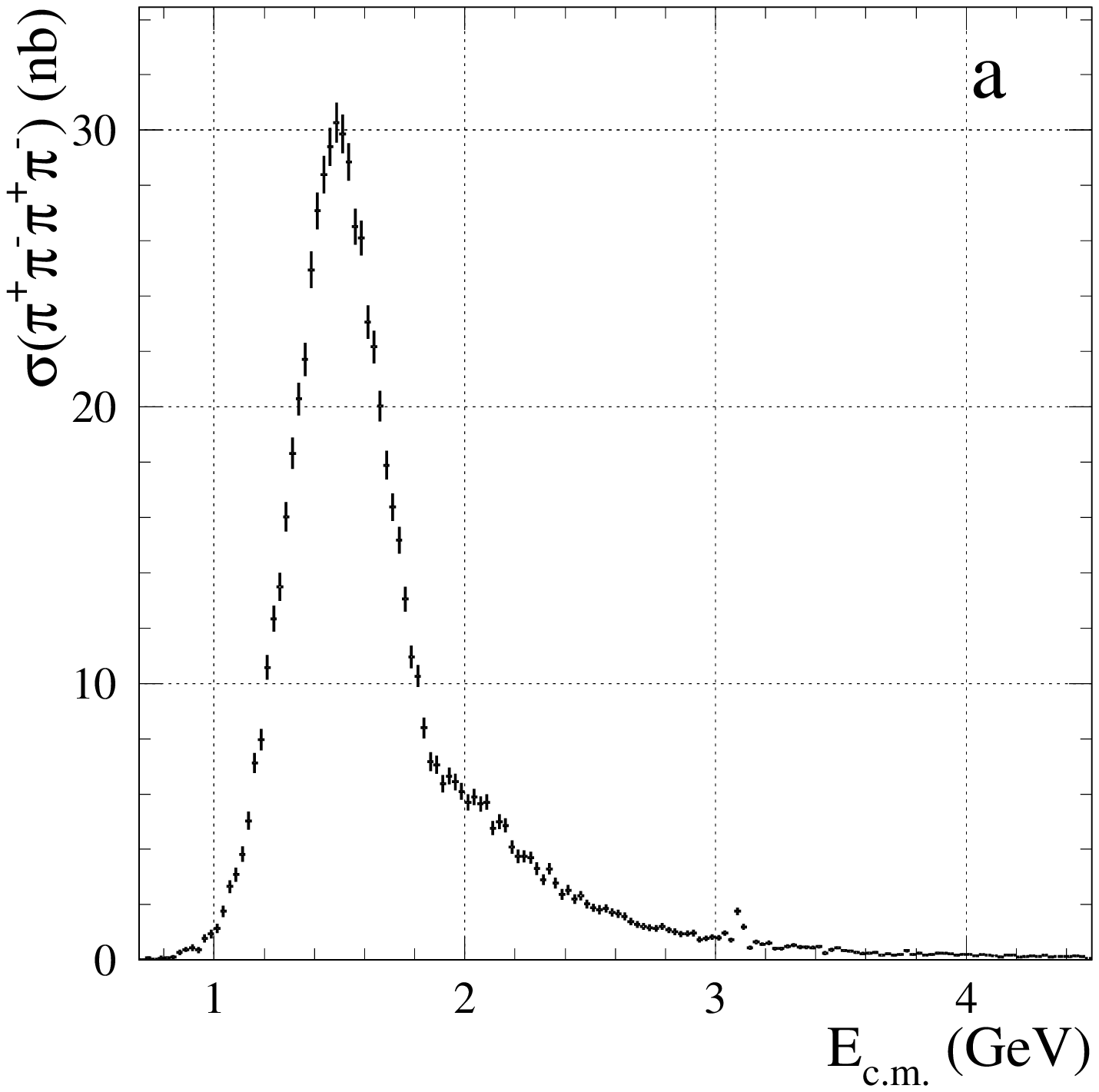}
\includegraphics[width=.32\linewidth]{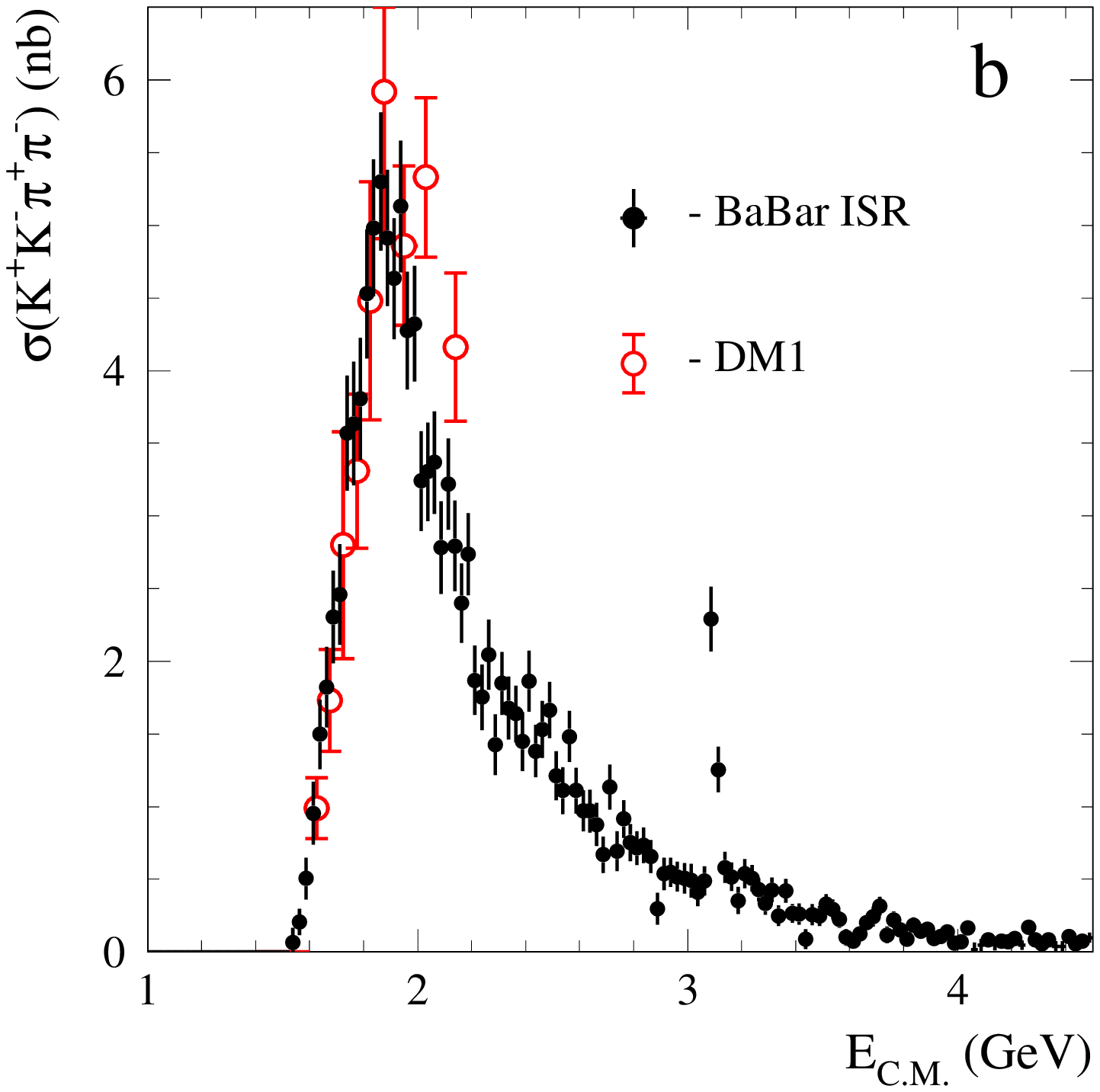}
\includegraphics[width=.32\linewidth]{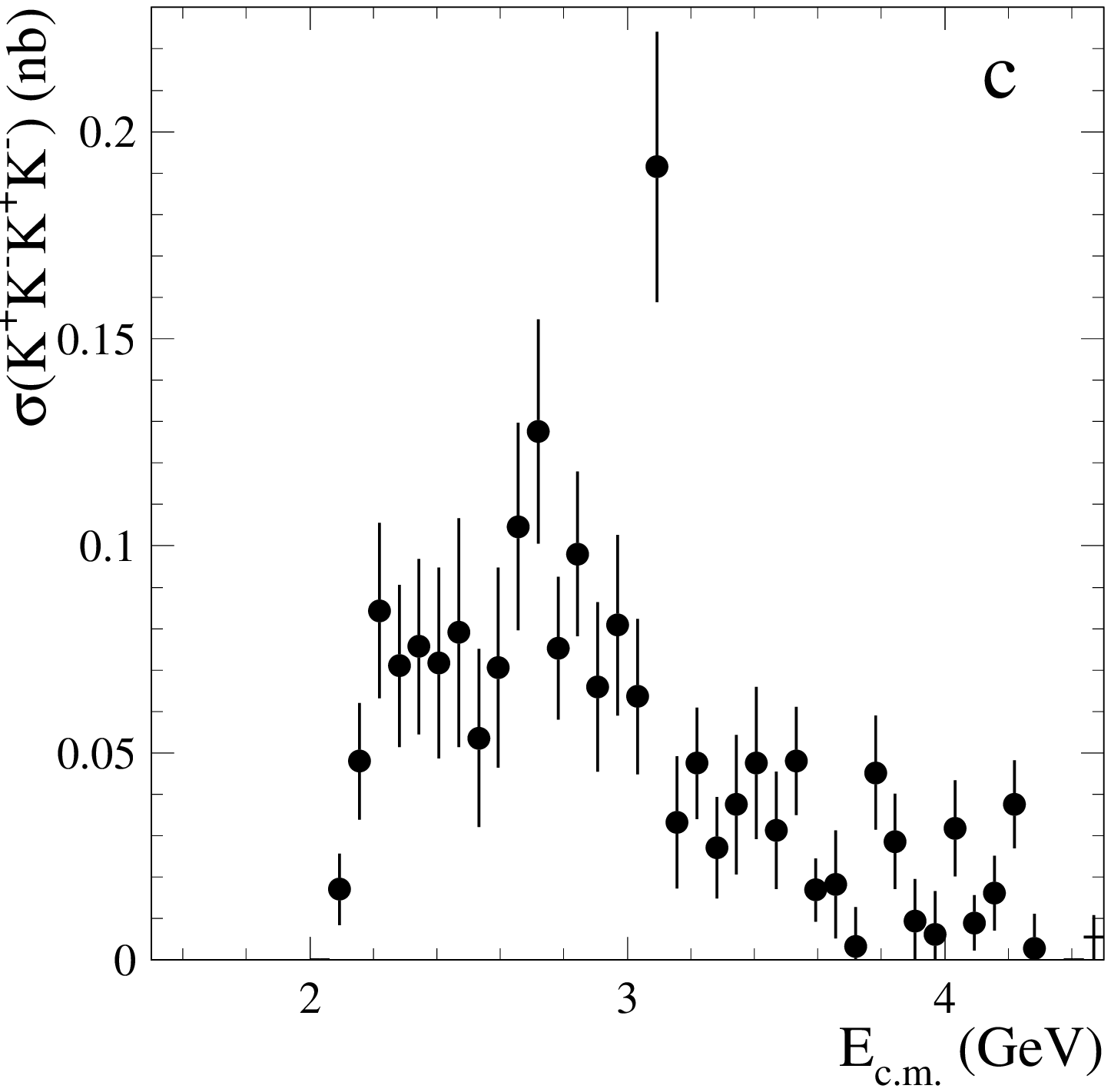}
\caption{The measured cross sections for $e^+e^-\to\pi^+\pi^-\pi^+\pi^-$ 
(a), $e^+e^-\to\pi^+\pi^-K^+K^-$ (b), $e^+e^-\to K^+K^-K^+K^-$
(c). \label{fig2}}
\end{figure}

\underline{\bf\boldmath The $\pi^+\pi^-\pi^+\pi^-,\,\pi^+\pi^-K^+K^-,\,
K^+K^-K^+K^-$ final states}~\cite{4pi}.
Our results on the measurement of the cross sections of $e^+e^-$ annihilation 
into four charged particles are shown in Fig.\ref{fig2}.
This $e^+e^-\to\pi^+\pi^-\pi^+\pi^-$ process ( Fig.\ref{fig2}a) has large 
cross section, and its accurate measurement is important for calculation of 
the hadronic  contribution into $(g-2)$ of muon. The BABAR data are in 
good agreement with previous direct $e^+e^-$ measurements. For energies 
above 1.4 GeV they have world's best accuracy.
From analysis of the 3-pion and 2-pion mass distributions 
we conclude that dominant intermediate state in this process is
$a_1(1260)\pi$. For energies above 2 GeV the $f_0(1370)\rho$ 
intermediate state is also seen.

The cross section for $e^+e^-\to K^+K^-\pi^+\pi^-$ is shown in
Fig.~\ref{fig2}b. This process proceeds via $K^\ast K\pi$, $\phi\pi\pi$,
and $\rho KK$ intermediate states. We also present the first measurement of
the $e^+e^-\to K^+K^-K^+K^-$ cross section (Fig.~\ref{fig2}c).

\underline{\bf\boldmath The $p\bar{p}$ final state}~\cite{ppbar}.
The cross section for $e^+e^- \to p\bar{p}$ depends on two form factors:
electric ($G_E$) and magnetic ($G_M$).
From measuring the total cross section we can extract a combination of
form factors. We define the effective form factor as 
$F(m)=\sqrt{(|G_M(m)|^2 + \tau |G_E(m)|^2)/(1+\tau)}$
where $\tau=2m_p^2/m^2$. Such definition allows to compare our
form-factor results with the data from previous $e^+e^-$ and $p\bar{p}$
experiments which use the assumption that $|G_E|=|G_M|$.
The ratio of the form factors can
be extracted from the analysis of the distribution of the
proton helicity angle ($\theta_p$) in the $p\bar{p}$ rest frame.
The ISR approach provides full $\theta_p$ coverage 
and hence high sensitivity to $|G_E/G_M|$. In contrast to previous 
$e^+e^-$ and $p\bar{p}$ experiments, our measurement of cross section 
does not use the assumption that $|G_E|=|G_M|$. It should be noted that 
$p\bar{p}$ mass resolution near the $p\bar{p}$ threshold is less than 
1 MeV and comparable with the energy spread of $e^+e^-$ machines.

\begin{figure}[h]
\includegraphics[width=.45\linewidth]{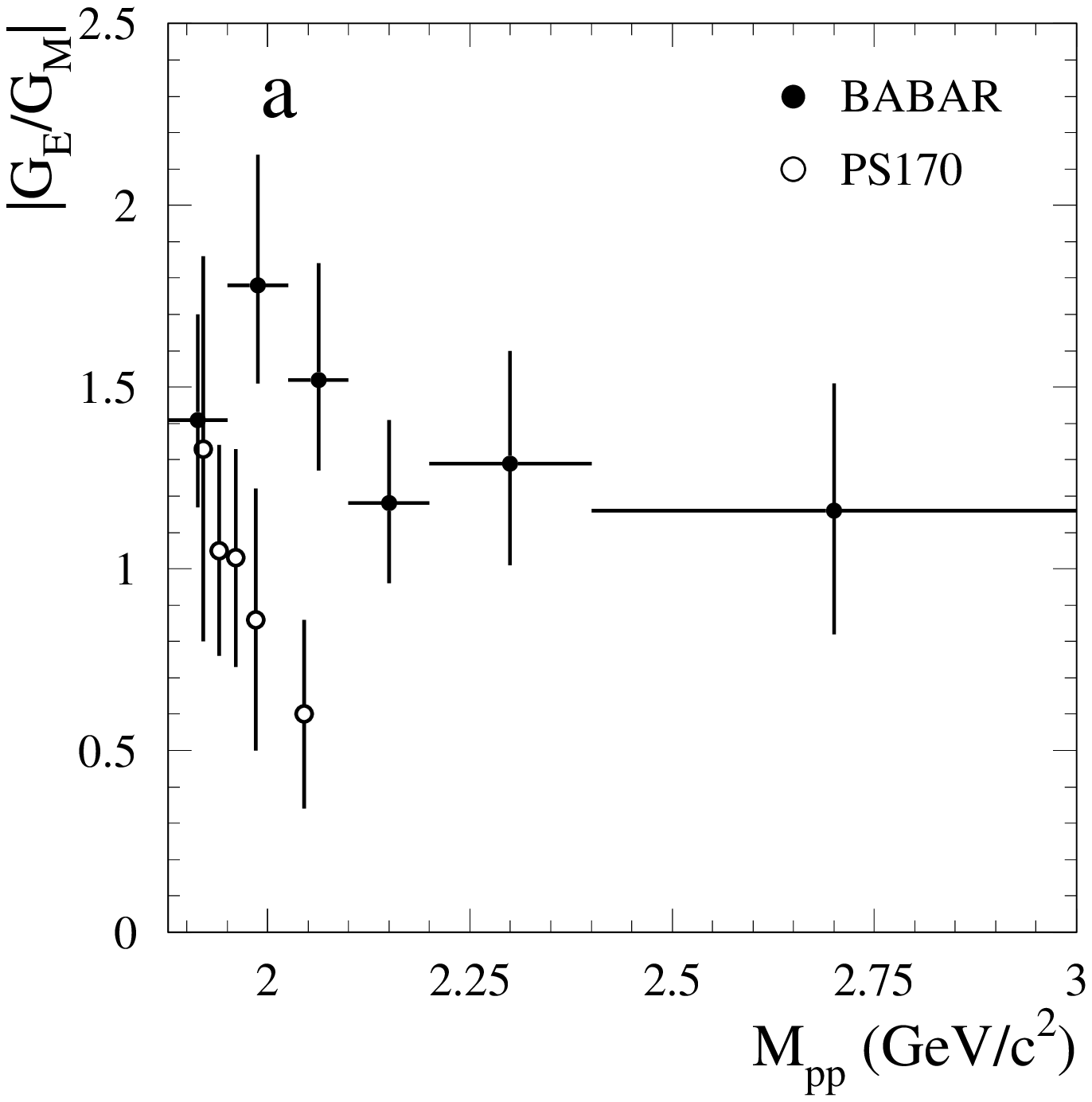}
\hfill
\includegraphics[width=.45\linewidth]{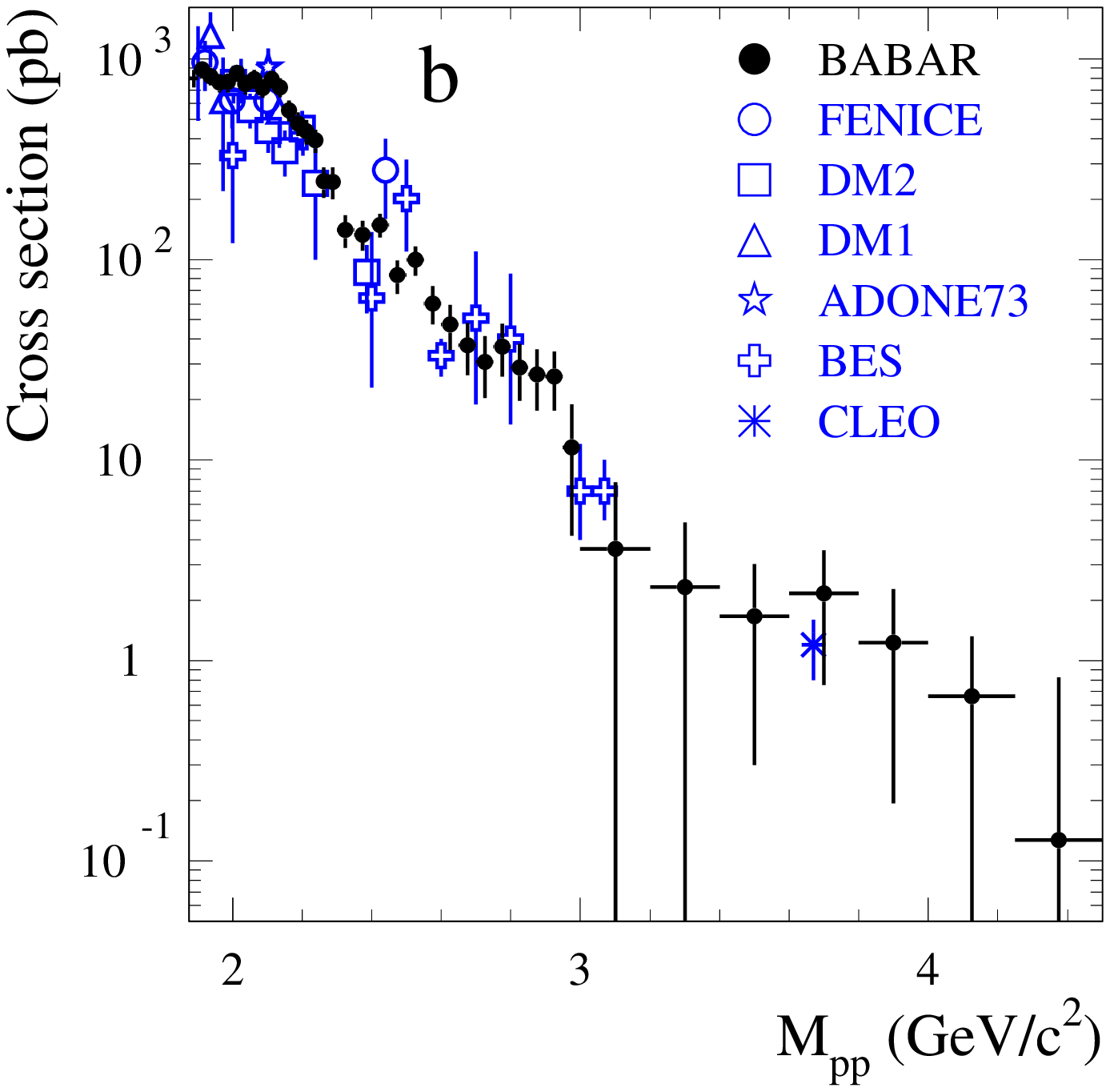}
\includegraphics[width=.45\linewidth]{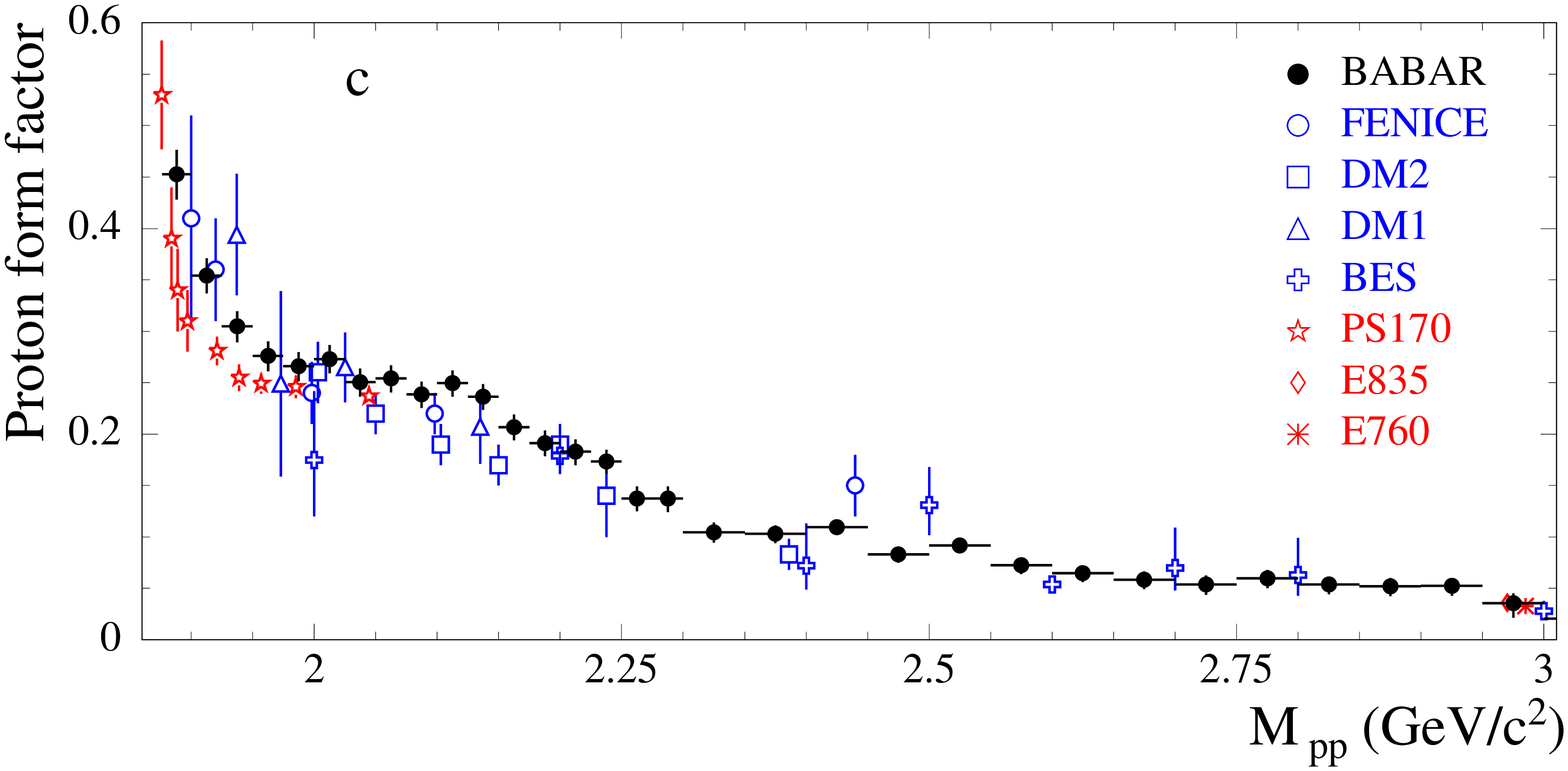}
\hfill
\includegraphics[width=.45\linewidth]{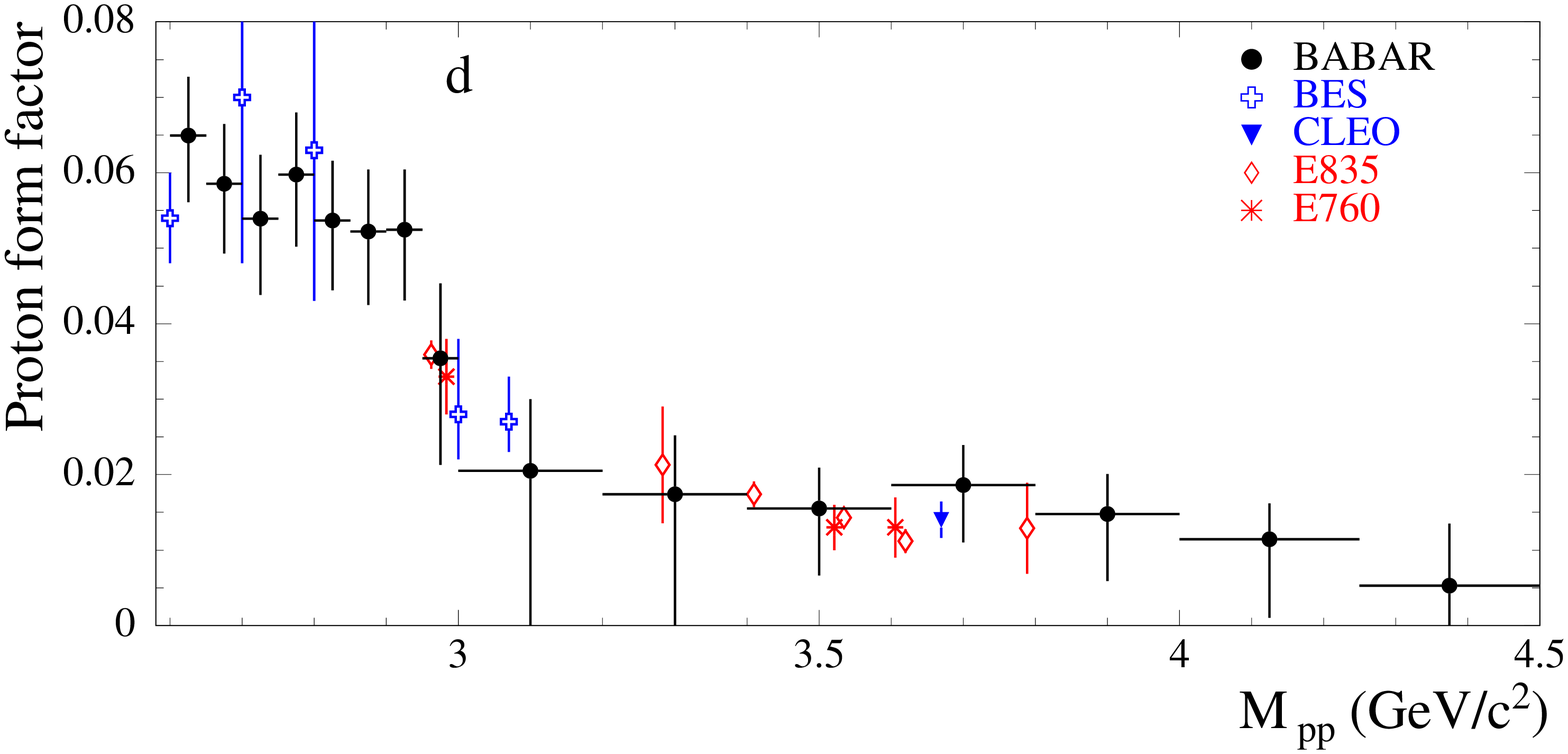}
\caption{a). Mass dependence
of the ratio $|G_E/G_M|$. b). The measured cross sections for
$e^+e^-\to p\bar{p}$.
c). and d).The effective proton form factor.
\label{fig3}}
\end{figure}

The $\cos\theta_p$ distribution was fitted in six bins of $m_{p\bar{p}}$
by a sum of two distributions corresponding $G_E$ and $G_M$
terms in the differential cross section. The $G_E$ distribution is
close to $\sin^2 \theta$, the  $G_M$ distribution to
$1+\cos^2 \theta$.
The obtained mass dependence of $|G_E/G_M|$ is shown in 
Fig.~\ref{fig3}a. Our result disagrees significantly
with previous measurement from LEAR~\cite{lear}.

The obtained  $e^+e^-\to p\bar{p}$ cross section shown in Fig.~\ref{fig3}b
(the contributions of $J/\psi$ and $\psi(2S)$ decays are subtracted)
is in reasonable agreement with previous $e^+e^-$ measurements. 
Our data more full and accurate for energy range below 3 GeV.
The effective form factor is shown in Fig.~\ref{fig3}c,d. It has a
complex mass dependence. We confirm the near-threshold enhancement
in the form factor observed in the LEAR experiment. 
There are also two mass regions, near 2.25 GeV/$c^2$ and 3 GeV/$c^2$,
that exhibit steep decreases in the form factor. This unusual
mass dependence has never been discussed in the literature. 

\begin{figure}
\includegraphics[width=.45\linewidth]{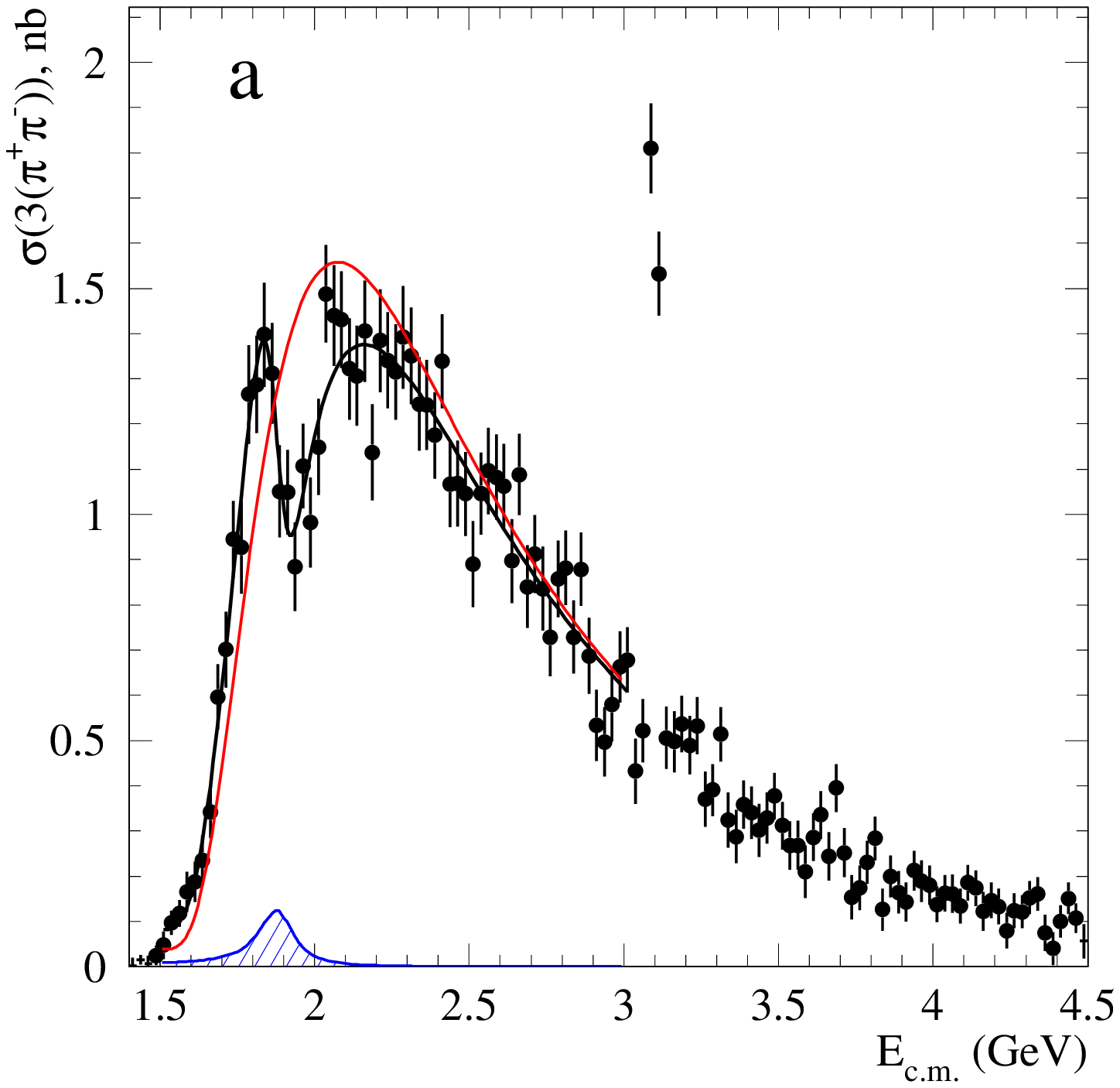}
\hfill
\includegraphics[width=.45\linewidth]{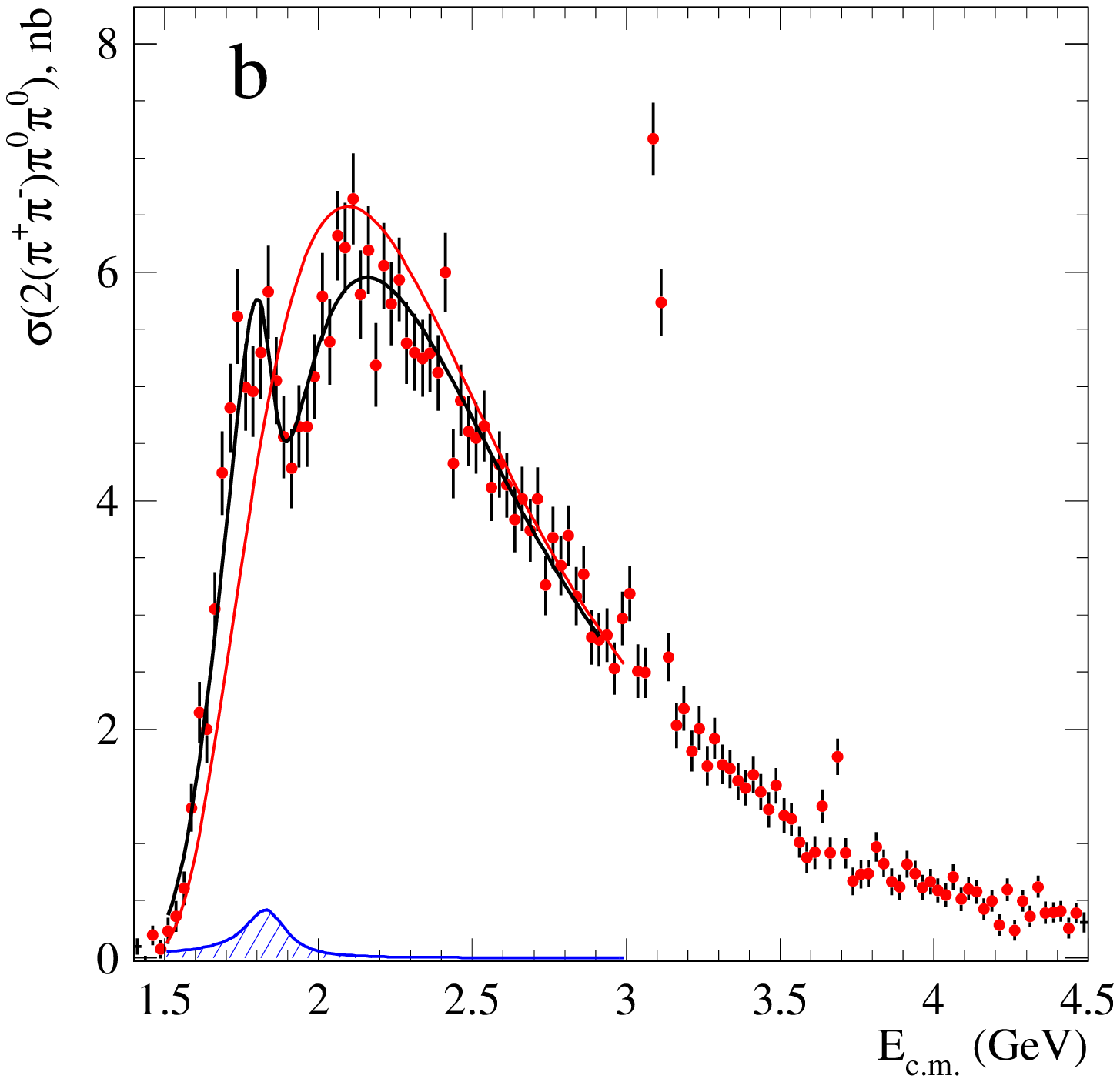}
\caption{The measured cross sections for $e^+e^-\to 3(\pi^+\pi^-)$ 
(a), $e^+e^-\to 2(\pi^+\pi^-\pi^0)$ (b).
\label{fig5}}
\end{figure}

\underline{\bf\boldmath The $6\pi$ final states}~\cite{6pi}.
The near-threshold enhancement in the $p\bar{p}$ mass spectrum 
may be a manifestation of $p\bar{p}$ subthreshold resonance. 
Such resonance may be seen in $e^+e^-\to hadrons$
cross section near 1.9 GeV. A narrow dip in the
$e^+e^-\to 6\pi$ cross section was observed in DM2 experiment
and in the diffractive photoproduction of six pions in FOCUS 
experiment~\cite{focus}. Our results on measurement of
six pion cross sections are shown in Fig.~\ref{fig5}.
In both six-pion cross sections the dips near 1.9 GeV are clearly seen. 
The cross sections are fitted by 
a sum of resonance and continuum amplitudes. The fit results 
the resonant mass $1880 \pm 30$ MeV/$c^2$ ($1860 \pm 20$  MeV/$c^2$)
and width $130 \pm 30$ MeV/$c^2$ ($160 \pm 20$ MeV/$c^2$)
for $3(\pi^+\pi^-)$ ($2(\pi^+\pi^-)2\pi^0$) mode.
The obtained width  is significantly differ from the 
FOCUS result of $29 \pm 14$ MeV/$c^2$.
\section{Summary}
The program to measure low-energy hadronic cross sections using
ISR is well underway at BaBar. We have presented the cross section 
measurements for 
$e^+e^-\to \pi^+\pi^-\pi^0,\, 2(\pi^+\pi^-),\, K^+K^-\pi^+\pi^-,\, 2(K^+K^-),\,
6\pi$. We significanly improved accuracy for
these cross sections in the 1.4--4.5 GeV energy range. 

We have also measured $e^+e^-\to p\bar{p}$ cross section and extracted
the proton effective form factor. These are most full and accurate 
$e^+e^-\to p\bar{p}$ data in the
energy range from the threshold up to 3 GeV. In contrast to previous experiment, our
measurement does not use the assumption that $|G_E|=|G_M|$. 
From analysys of the proton angular distribution we have extracted
the energy dependence of the $|G_E/G_M|$ ratio. Its value is found to be 
significanly greater than unity for energies below 2.1 GeV. 

The analysys of $\pi^+\pi^-$, $K^+K^-$, $K\bar{K}\pi$, $\,\pi^+\pi^-\pi^0\pi^0$, $5\pi$,
and $\,D^{(*)}\bar{D}^{(*)}$ final states are in progress at BaBar.

\end{document}